# C and Si allotropes and derived SiC from first principles.

Samir F. Matar[*]

Lebanese German University (LGU), Sahel-Alma, Jounieh, Lebanon.
https://orcid.org/0000-0001-5419-358X

Author emails: abouliess@gmail.com

## Abstract

Novel extended networks of $C_8$, $Si_8$ and silicon carbide $Si_4C_4$ are proposed based on crystal chemistry rationale and optimized structures to ground state energies and derived physical properties within the density functional theory (DFT). The two carbon and silicon allotropes and the silicon carbide belong to primitive tetragonal space group $P\bar{4}m2$ N°115. $C_8$ allotrope structure made of corner sharing *C4* and *Si4* tetrahedra is illustrated by charge density projections exhibiting $sp^3$-like carbon hybridization. From careful symmetry analysis, $C_8$ is found as another representation of cubic diamond. It is identified as ultra-hard with a similar magnitude of Vickers hardness. The interest in $C_8$ is to serve as template to study $Si_8$ and Si-C binary. $Si_8$ allotrope is found soft with $H_V = 13$ GPa alike cubic Si, and $Si_4C_4$ is identified with $H_V = 33$ GPa close to experimental SiC. All three new phases are mechanically (elastic constants) and dynamically (phonons) stable, and their electronic band structures are characteristic of insulating $C_8$ (diamond) with large band gaps of about 5 eV, and semi-conducting $Si_8$ and $Si_4C_4$ with band gaps of about 1 eV.

# 1. Introduction and context

The chemistry of carbon and silicon, major elements in nature, is ruled by three several factors mainly involving:

(i)     the difference of –Pauling– electronegativities with $\chi(C) = 2.55 > \chi(Si) = 1.80$ leading to polar Si–C bonds with a trend of charge transfer Si $\rightarrow$ C .

(ii)    the larger size of Si covalent radius with respect to C: $rSi = 1.14$ Å versus $rC = 0.76$ Å resulting from one extra shell for Si: Si $(1s^2, 2s^2, 2p^6, 3s^2, 3p^2)$ versus C $(1s^2, 2s^2, 2p^2)$ leading to more compressible Si and Si based compounds as SiC examined herein.

(iii)   the tetrahedral coordination (sp3 -like), common to Si, C,

In this context, diamond is the hardest material while isostructural silicon is soft and binary combinations between Si and C, SiC, shows intermediate hardness and considered as an abrasive material [1]. Regarding the valence electron count C, Si and SiC are isoelectronic with integer multiple of 4: C (4 electrons), Si (4 electrons) and SiC (2×4 electrons) and expected to have similar electronic band structures i.e., insulating like diamond and semiconducting like Si and SiC. Structurally, the 3D ly the diamond cubic structure (in primitive cubic cell with 2x4=8 atoms) with the arrangement of corner sharing *C4* tetrahedra where the angle $\angle$C-C-C =109.47° highlight the C sp³-like hybridization.

In last decades large research efforts were devoted to identifying novel allotropes of carbon close to diamond using modern materials research software as CALYPSO [3] and USPEX [4]. Despite their success, thorough investigations based on energy criteria are more accurately achieved thanks to calculations based on the quantum mechanics Density Functional Theory (DFT) framework devised in two papers: in 1964 for the theoretical framework by Hohenberg and Kohn [5] and followed in 1965 by Kohn and Sham [6] who established the Kohn-Sham equations for solving the wave equation practically in calculation codes built around the DFT (cf. next section).

Such investigations, backed with liminary crystal chemistry rationale led to several works by us and others on novel carbon allotropes with different stoichiometries as hexacarbon $C_6$ [7] and rhombohedral rh-$C_4$ (or in hexagonal axes: h-$C_{12}$) [8] that is another representation of diamond, both 3D carbon allotrope being close to diamond in their physical (Vickers hardness

$H_V \sim 100$ GPa) and dynamical (phonons, heat capacity, entropy), as well as electronic band structure properties.

Extended carbon networks [9] were investigated to address the challenging question: "*is diamond the hardest material*" as generally admitted? In this context, super-cubane $C_8$ was announced by Johnston and Hoffmann as super-dense allotrope [10]. Like in former works [2, 7, 8], the present study focuses on novel $C_8$ 3D allotrope based on crystal chemistry rationale with geometry optimized structures to the ground state within DFT-based calculations. The results highlight cohesive properties as well as ultra-hardness and related electronic structure properties like diamond, noting that $C_8$ is actually another view of cubic diamond. More specifically, the mechanical (elastic constants) and dynamical (phonons bands) stabilities of the novel octacarbon allotropes were confirmed and identified with thermal properties (heat capacity) alike diamond. The electronic band structures are shown to exhibit insulating behavior like diamond with large band gap of 5 eV.

## 2. Computational framework

For the search for the ground state structures, geometry optimizations calculations onto the ground state with minimal energies were performed using DFT-based plane-wave Vienna Ab initio Simulation Package (VASP) [11,12]. For the carbon atomic potential including valence states, the projector augmented wave (PAW) method was applied [12, 13]. Treating at the same level the exchange X and the correlation C, the exchange-correlation (XC) effects were considered using a generalized gradient approximation (GGA) [14]. The relaxation of the atoms onto ground state geometry was done applying a conjugate-gradient algorithm [15]. Blöchl tetrahedron method [16] with corrections according to Methfessel-Paxton scheme [17] was applied for geometry optimization and energy calculations. A special *k*-point sampling [18] was applied for approximating the reciprocal space Brillouin-zone (BZ) integrals. For better reliability, the optimization of the structural parameters was carried out along with successive self-consistent cycles with increasing mesh until the forces on atoms were less than 0.02 eV/Å and the stress components below 0.003 eV/Å$^3$.

Besides the elastic constants calculated to infer the mechanical stabilities and hardness, further calculations of phonon dispersion curves were also carried out to verify the dynamic stability of the new carbon allotropes. In the present work, the phonon modes were computed considering the harmonic approximation via finite displacements of the atoms around their equilibrium positions to obtain the forces from the summation over the different configurations. The phonon dispersion curves along the direction of the Brillouin zone are subsequently obtained using "Phonopy" interface code based on Python language [19]. Thermodynamic properties such as the heat capacity $C_v$ and the entropy S, were calculated as functions of temperature.

## 3. Results and discussions

### 3-1-Energy and crystal symmetry

Quite recently, we proposed a novel tetracarbon as the simplest 'seed' of corner sharing *C4* tetrahedra for building large carbon edifices approaching diamond [20]. An original $C_8$ structure was built starting from such simple unit leading to a tetragonal lattice that was fully geometry optimized to stress-free ground state. As a first assessment, energy is the prevailing criterion. The total energies: $E_{total}$(eV) and the derived atom-averaged energy $E_{atom}$(eV) are presented in Table 1 with those obtained for Diamond considered in simple cubic with 8 atoms/cell.

Strikingly, tetragonal $C_8$ is found energetically very close to diamond. The densities of both systems are also very close.

The crystal structures are shown in Fig. 1a and 1b showing the corner sharing tetrahedral 3D carbon networks. The angle ∠C-C-C =109.47° and d(C-C) = 1.54 Å are found in both structures thus providing signatures of purely covalent 3D carbon networks.

Table 2 details the crystal data of diamond and the new octacarbon. For diamond, a good agreement is found between calculated and literature cubic lattice constant [21]. Interestingly, while carbon is identified in a single Wyckoff position at 8*a* (0,0,0) in space group $Fd\overline{3}m$, (N° 227), tetragonal $C_8$, in space group $P\overline{4}m2$ (N° 115) shows 4 different single and double occupation carbon positions. Such a feature is relevant as template to model specific substitutions in diamond while keeping high symmetry as with B (p-doping) or N (n-doping) in view of tuning the physical properties of diamond.

### 3-2. Charge density 3D projections

The "electronic ↔ crystal structure" relationship is further illustrated with the charge density projections. Fig. 2 shows the charge density volumes (yellow) with perfect $sp^3$ tetrahedral shape around carbon in the tetragonal and hexagonal systems as equally within diamond. Then tetrahedral $C(sp^3)$-like carbon is the building unit.

### 3-3. Mechanical properties

(i)     *Elastic constants*

The investigation of mechanical characteristics was based on the calculations of the elastic properties determined by performing finite distortions of the lattice and deriving the elastic constants from the strain-stress relationship. Most compounds are polycrystalline, and generally considered as randomly oriented single crystalline grains. Consequently, on a large scale, such materials can be considered as statistically isotropic. They are then fully described by bulk ($B$) and shear ($G$) moduli obtained by averaging the single-crystal elastic constants. The method used here is Voigt's [22], based on a uniform strain. The calculated sets of elastic constants are given in Table 3. All values are positive. Their combinations obeying the rules pertaining to the mechanical stability of the phase, and the equations providing the bulk $B_V$ and shear $G_V$ moduli are as follows for the tetragonal system [23]:

$C_{ii}$ (i =1, 3, 4, 6) > 0; $C_{11} > C_{12}$, $C_{11} + C_{33} - 2C_{13} > 0$; and $2C_{11} + C_{33} + 2C_{12} + 4C_{13} > 0$.

$B_{Voigt}^{tetr.} = 1/9$ ($2C_{11} + C_{33} + 2C_{12} + 4C_{13}$); and $G_{Voigt}^{tetr.} = 1/15$ ($2C_{11} + C_{12} + 2\,C_{33} - 2C_{13} + 6C_{44} + 3C_{66}$).

The calculated values for diamond and tetragonal $C_8$ are given in Table 3. Both exhibit large $B_V$ (> 400 GPa) and $G_V$ (> 500 GPa) magnitudes that are comparable with the accepted values for diamond $B_V$ =445 GPa and $G_V$ = 550 GPa [21]. Not that the bulk modulus is equal to the value obtained for tet.$C_4$ [20] but the novel tet.$C_8$ structure with the original stacking of *C4* tetrahedra offers a 9 GPa larger shear modulus of 583 GPa versus $G_V$(tet.$C_4$)= 574. The consequence is that a larger hardness is expected for tet.$C_8$.

(ii) *Hardness*

Vickers hardness ($H_V$) was predicted using four pertinent theoretical models of hardness [24-27]. The thermodynamic model [24] is based on thermodynamic properties and crystal structure, while Mazhnik-Oganov [25] and Chen-Niu [26] models use the elastic properties. Lyakhov-Oganov approach [27] considers topology of the crystal structure, strength of covalent bonding, degree of ionicity and directionality. The fracture toughness ($K_{Ic}$) was evaluated within the Mazhnik-Oganov model [25].

The results are summarized in Table 4 presenting Vickers hardness calculated using the different theoretical models and other mechanical properties such as shear modulus ($G$), Young's modulus ($E$), the Poisson's ratio ($\nu$) and fracture toughness ($K_{Ic}$). Interestingly the hardness of the tet.$C_8$ shows larger magnitude than diamond with Mazhnik-Oganov [25] model and Chen at al. model [26] but not with the thermodynamic model [24] where $H_v$, is found equal to the value admitted for diamond. However, concomitantly with Mazhnik-Oganov model, the fracture toughness: $K_{Ic}$(tet.$C_8$) = 6.7 MPa·m$^{½}$ is found larger than calculated for diamond with $K_{Ic}$ = 6.4 MPa·m$^{½}$. Consequently, the novel octacarbon allotrope can be considered as prospective ultra-hard material, at least as good as diamond.

*3-4. Dynamical stabilities from the phonons.*

Beside structural stability criteria observed for the new carbon allotrope from the positive magnitudes of the elastic constants and their combinations, the phonon modes were subsequently computed. Phonons are quanta of vibrations; their energy is quantized through the Planck constant 'h' used in its reduced form ℏ (ℏ = h/2π) giving with the wave number ω the energy: E = ℏω.

Besides the novel allotrope the phonon band structures were obtained for diamond for the sake of comparison. Fig. 3 shows the phonon bands. Along the horizontal direction, the bands run along the main lines of the respective Brillouin zone (reciprocal k- space), i.e., cubic (Fig. 3a) and tetragonal (Fig. 3b). The vertical direction shows the frequencies given in units of terahertz (THz). Since no negative frequency magnitudes are observed, expectedly for diamond, but also for tetragonal $C_8$, the structure can be considered as dynamically stable. There are 3N-3 optical modes at higher energy than three acoustic modes starting from zero energy (ω = 0) at the Γ point, center of the Brillouin Zone, up to a few Terahertz. They correspond to the lattice rigid translation modes of the crystal (two transverse and one longitudinal). The remaining bands are 21 but the higher the symmetry, the degeneracy is observed with bands found at a given frequency, and a reduction in the number of dispersion curves is observed as obvious from the comparison of diamond few bands and the larger number of bands for the two octacarbon polymorphs. In the three panels the energy range is the same for all three phases, i.e. from 0 to 40 THz, stressing furthermore their similar behaviors, especially with the magnitude observed for diamond by Raman spectroscopy: $\omega \sim 40$ THz [28].

*3-5. Thermal properties.*

Following the phonon band structures, the thermal properties such as the entropy and the heat capacity $C_v$ were calculated using the statistical thermodynamic expressions from the phonon frequencies on a high precision sampling mesh in the BZ (cf. the textbook by Dove on 'Lattice Dynamics' [26]). As shown in Fig. 4, tetragonal $C_8$ and Diamond present similar curves for the temperature change of the entropy and heat capacity at constant volume. The black curve stands for the temperature change of Helmholtz free energy: $F = U - TS$ where U stands for the internal energy and S for entropy. For the two carbon allotropes, the free energy decreases with temperature as expected from the equation above because S increases with T almost linearly as it can be seen in both panels. The entropy S and the heat capacity are close to zero up to 100 K. Above 100 K, S increases continuously and almost linearly up to the highest temperatures. For $C_v$ a validation of the calculation results was found from experimental data on diamond up to high temperatures obtained by Victor back in 1962 [27]. The discrete experimental points obtained from 300K up to 1000K by steps of 100K are plotted as blue diamond symbols on the calculated curve (green). They are found exactly on the calculated diamond curve (Fig. 4a), and only slightly below the calculated $C_V$ curve in *tet*.$C_8$ (Fig. 4a), thus providing an additional proof for the closeness of the new carbon allotrope to diamond.

*3.6 Electronic band structures.*

Fig. 5 shows the electronic band structures obtained using the all-electrons DFT-based augmented spherical method (ASW) [28]. For the sake of comparison, the band structure of diamond is exhibited. The bands develop along the main directions of the cubic (diamond) and tetragonal (tet.$C_8$) Brillouin zones,

The energy level along the vertical line is with respect to the top of the valence band (VB), $E_V$. As a specific character of diamond (Fig. 5a), the band gap is indirect with a magnitude slightly larger than 5 eV, whereas at the center of the Brillouin zone the energy gap is slightly larger, as observed in the literature [29]. This feature also characterizes the tetragonal allotrope exhibiting indirect band gap between $\Gamma_{VB}$ and $Z_{CB}$., i.e., along $k_z$ .

## 4. Conclusions

In this work a novel ultra-hard carbon allotrope, $tet$.$C_8$ that has be regarded as another representation of diamond, was proposed with electronic, mechanical, dynamic, and thermal properties close to diamond. The hardness assessment provides Vickers hardness hypothesized larger than diamond along different models of approaching hardness.

Table 1  Crystal parameters of tetragonal with space group $P\overline{4}m2$ N°115: $C_8$, $Si_8$, and $Si_4C_4$ from DFT calculations. Lattice constants and interatomic distances are units of Å (1 Å = $10^{-10}$ m).

a)  $C_8$ and $Si_8$ (Fig. 1a & 1c).

$C_8$ : $a = b = 2.519$; $c = 7.126$
$Si_8$ : $a = b = 3.864$; $c = 10.933$

| $C_8$ | $Si_8$ | Wyckoff | x | y | z |
|---|---|---|---|---|---|
| C1 | Si1 | 1a | 0.0 | 0.0 | 0.0 |
| C2 | Si2 | 1d | 0.0 | 0.0 | ½ |
| C3 | Si3 | 2f | ½ | ½ | ¼ |
| C4 | Si4 | 2g | ½ | 0.0 | 1/8 |
| C5 | Si5 | 2g | ½ | 0.0 | 5/8 |

d(C-C) = 1.54
d(Si-Si) = 2.37

b)  $Si_4C_4$ (Fig. 1b).
$a = b = 3.090$; $c = 8.748$.

| Atom | Wyckoff | x | y | z |
|---|---|---|---|---|
| C1 | 1a | 0.0 | 0.0 | 0.0 |
| C2 | 1d | 0.0 | 0.0 | ½ |
| C3 | 2f | ½ | ½ | ¼ |
| Si1 | 2g | ½ | 0.0 | 1/8 |
| Si2 | 2g | ½ | 0.0 | 5/8 |

d(Si-C) = 1.893.

Table 2   Elastic constants $C_{ij}$ and Voigt values of bulk ($B_V$) and shear ($G_V$) moduli and Vickers hardness ($H_V$). All values are in GPa.

| | $C_{11}$ | $C_{12}$ | $C_{13}$ | $C_{33}$ | $C_{44}$ | $C_{66}$ | $B_V$ | $G_V$ | $H_V$ |
|---|---|---|---|---|---|---|---|---|---|
| $C_8$ | 1171 | 29 | 133 | 1067 | 465 | 567 | 444 | 582 | 113 |
| $Si_4C_4$ | 510 | 37 | 162 | 422 | 127 | 246 | 241 | 208 | 34 |
| $Si_8$ | 180 | 32 | 58 | 153 | 48 | 73 | 90 | 65 | 13 |

\*$H_V = 0.92 \ (G/B)^{1.137} \ G^{0.708}$

See SCON PAPER

Hv(Si) wiki➔ 11GPa

SiC >> 41 GPa

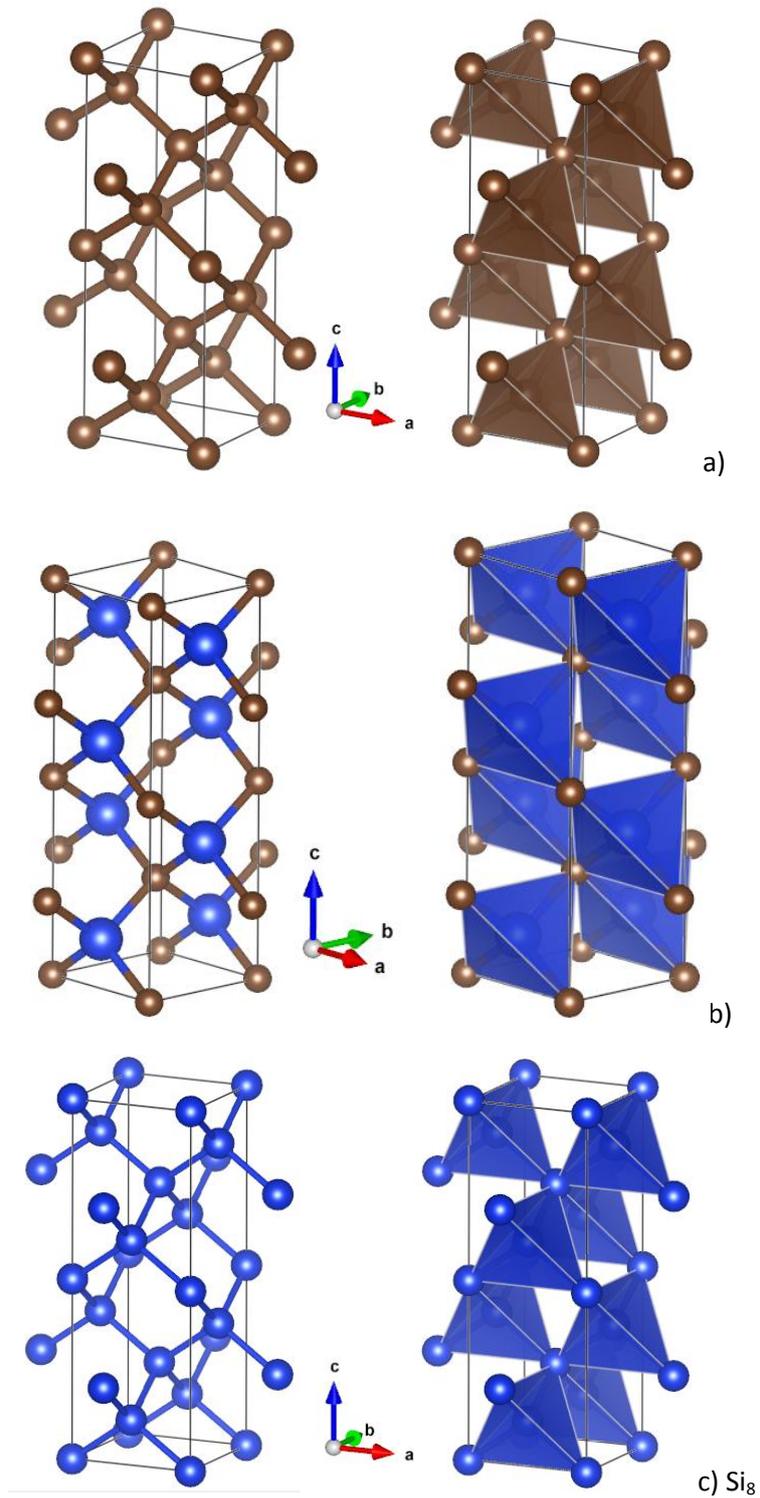

Fig. 1- Crystal structures of the tetragonal phases. a) $C_8$, b) $Si_4C_4$ c) $Si_8$ highlighting the corner-sharing tetrahedral 3D arrangement. Brown and blues spheres represent carbon and silicon respectively.

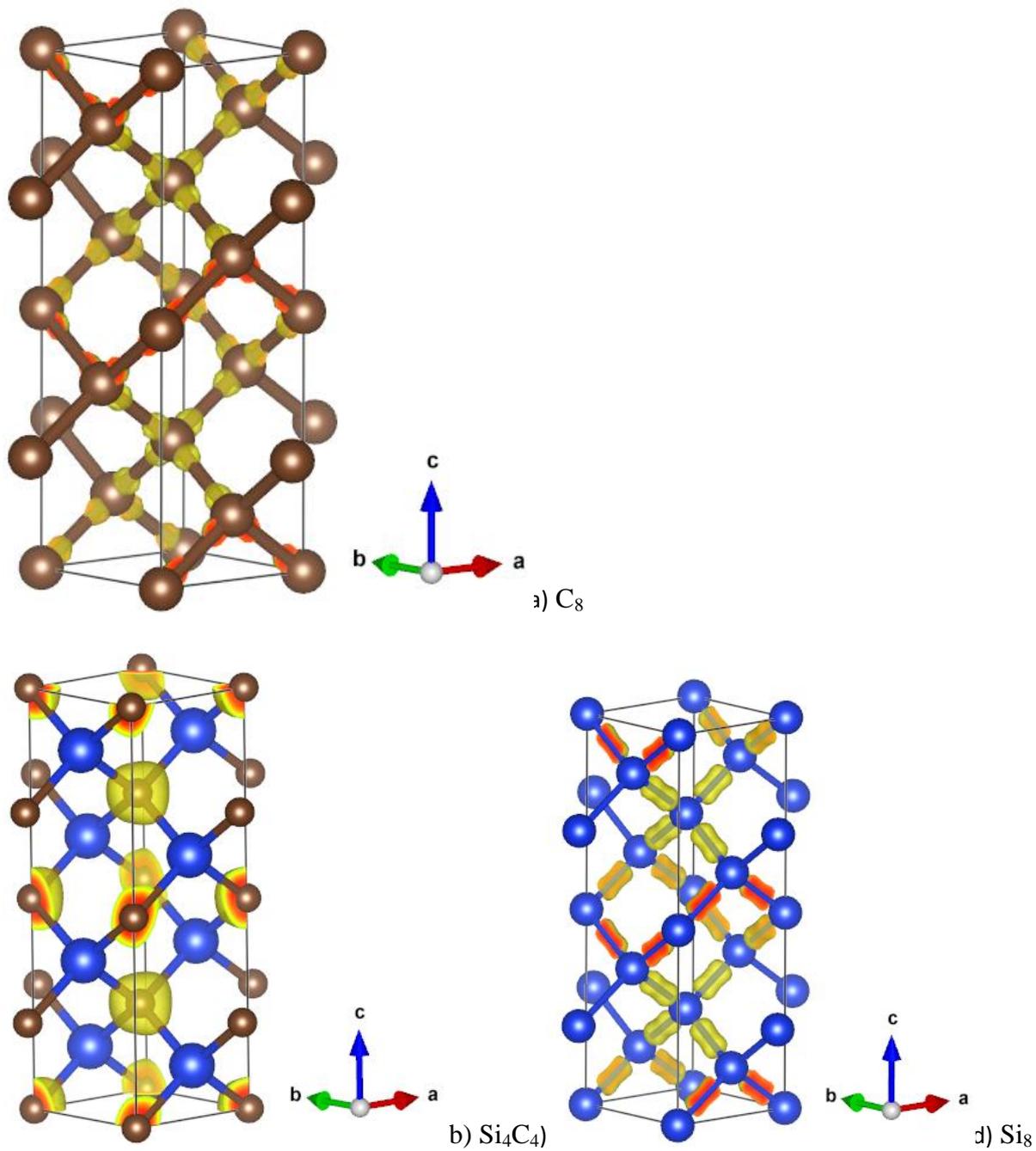

Fig. 2. Charge density projections (yellow volumes) of the tetragonal phases. a) $C_8$, b) $Si_4C_4$, and c) $Si_8$. Brown and blues spheres represent carbon and silicon respectively.

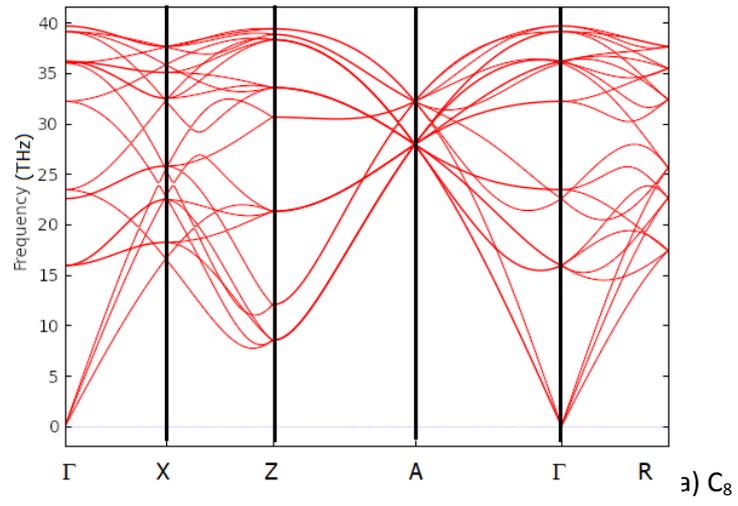

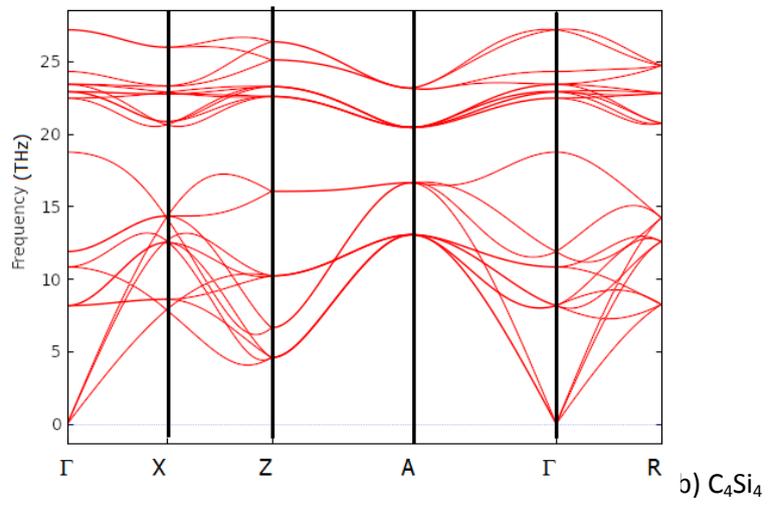

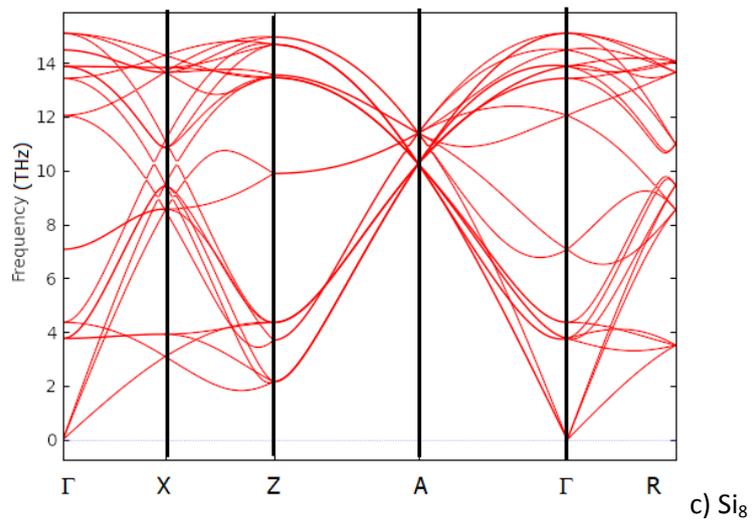

Fig. 3. Phonons band structures of a) $C_8$, b) $Si_4C_4$, and c) $Si_8$.

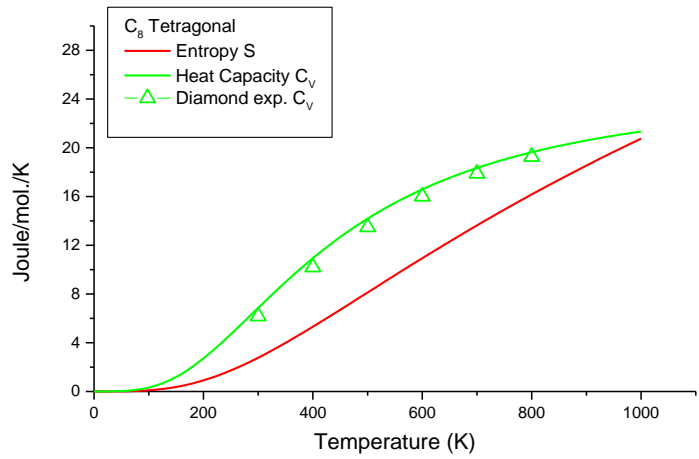

a) C$_8$

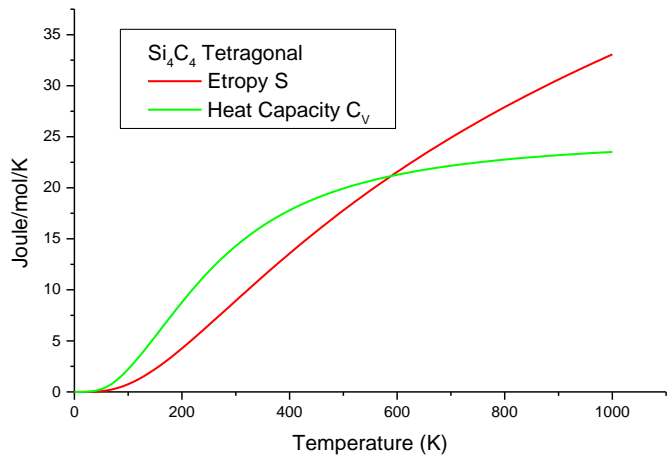

b) C$_4$Si$_4$

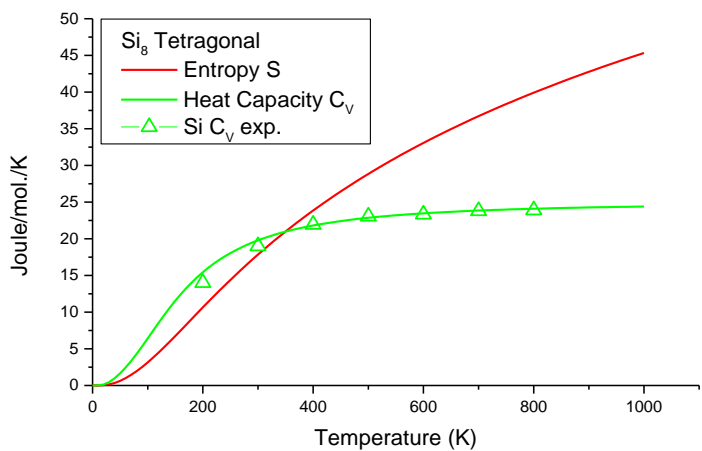

c) Si$_8$

Fig. 4. Thermal properties, entropy S and specific heat C$_V$ as functions of temperature of the three tetragonal phases: a) C$_8$ with the experimental C$_V$ values from Ref. [38] shown as open symbols; b) Si$_4$C$_4$; c) Si$_8$ with experimental C$_v$ values from Ref. [XX] shown as open symbols.

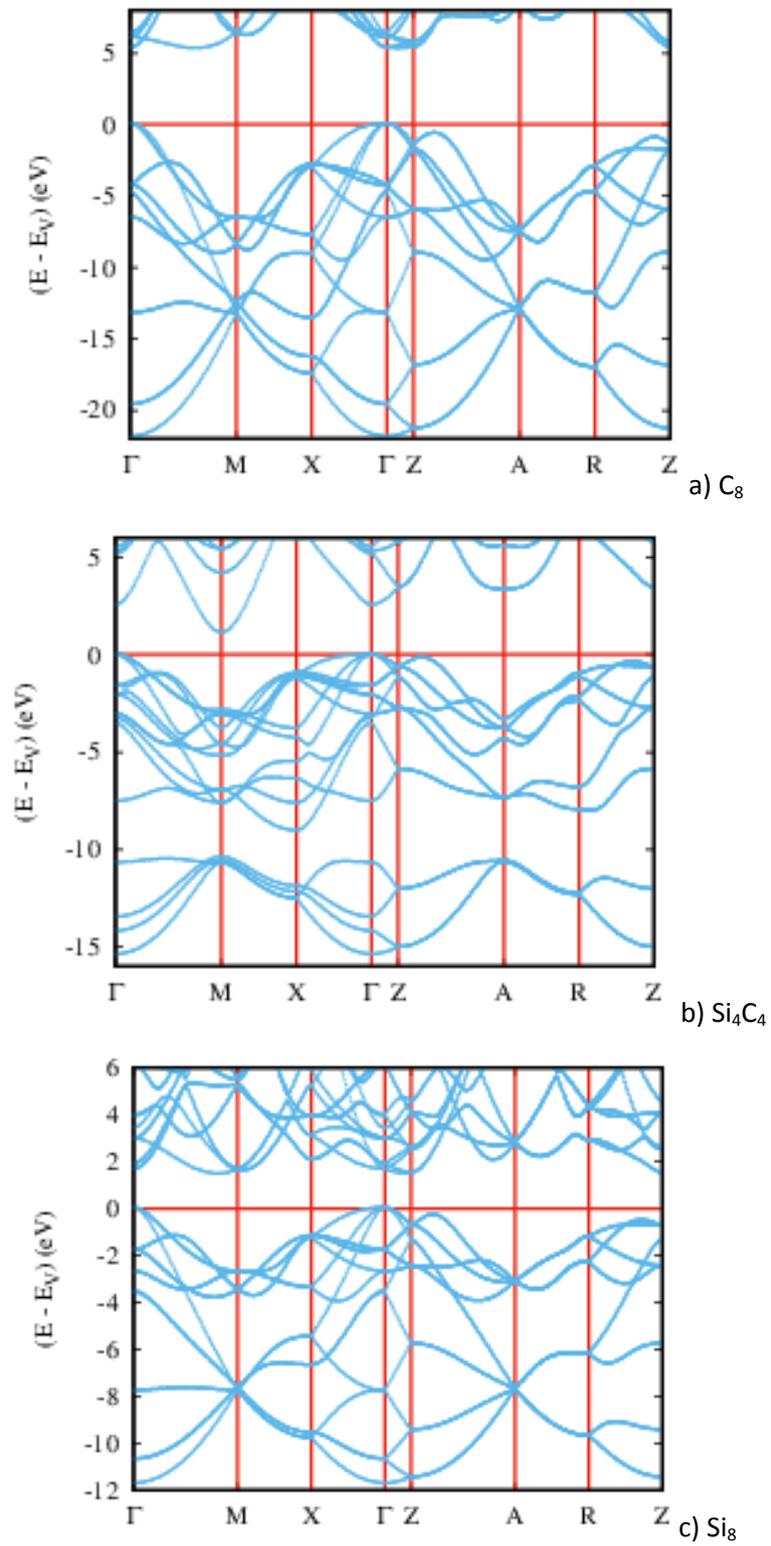

Fig. 5. Electronic band structures along the major lines of the tetragonal Brillouin zone.
a) $C_8$, b) $Si_4C_4$, and c) $Si_8$